\documentclass{article}
\usepackage[LGR,T1]{fontenc}
\usepackage[utf8]{inputenc}
\usepackage{bm}
\usepackage{amsmath}
\usepackage{amsthm}
\usepackage{amssymb}
\usepackage{graphicx}
\usepackage{esint}

\makeatletter

\ProvideTextCommand{\~}{LGR}[1]{\char126#1}

\newcommand{\lyxaddress}[1]{
	\par {\raggedright #1
	\vspace{1.4em}
	\noindent\par}
}

\@ifundefined{date}{}{\date{}}
\usepackage{tikz}
\usepackage{fancyhdr}
\fancypagestyle{plain}{
  \fancyhf{}
  \fancyhead[L]{OUJ-FTC-13}
  
}

\usepackage[english]{babel}

\@ifundefined{showcaptionsetup}{}{%
 \PassOptionsToPackage{caption=false}{subfig}}
\usepackage{subfig}
\makeatother

\begin{document}
\title{Holography of Transmission Lines: Insights of Continuous MERA and
AdS/CFT}
\author{So Katagiri\thanks{So.Katagiri@gmail.com}}
\maketitle

\lyxaddress{\textit{Nature and Environment, Faculty of Liberal Arts, The Open
University of Japan, Chiba 261-8586, Japan}}
\begin{abstract}
This study examines the holographic representation of the quantum
theory of transmission lines, which play a crucial role in quantum
computing and quantum information. Utilizing Yurke and Denker's quantum
circuit network theory within the framework of continuous MERA (cMERA)
in AdS space, we analyze the quantization and interactions of transmission
lines. The metric is revealed to be described by the inductance of
the quantum circuit, which is AdS-space in its 0-limit. These results
provide new insights into handling and controlling complex phenomena
in quantum circuits, potentially advancing the understanding of quantum
computing and quantum communication. 
\end{abstract}

\section{Introduction}

Transmission lines, fundamental in classical electromagnetism for
signal transmission, have evolved significantly with the advent of
quantum technologies. Historically rooted in classical theory, the
understanding of transmission lines has been recontextualized through
the lens of quantum mechanics\cite{Yurke_1984}, particularly with
the discovery of the Josephson junction in the late 1980s\cite{Clarke_1988,Devoret_1985,Martinis_1985}.
This breakthrough paved the way for the development of superconducting
circuits, a cornerstone in the quest for scalable quantum computing.
The late 1990s saw the experimental validation of superconducting
qubits, marking a pivotal moment in quantum information science\cite{Nakamura_1999}.
These qubits exhibited coherent quantum oscillations, with subsequent
advancements leading to significant improvements in coherence times
as qubit designs and couplings were refined \cite{Pashkin_2003}\cite{Yamamoto_2003}.
In the early 2000s, the principles of cavity quantum electrodynamics
(Cavity QED), traditionally developed in atomic physics and quantum
optics, were adapted to superconducting circuits\cite{Haroche_2006}\cite{Kimble_1998}.
This adaptation gave rise to circuit quantum electrodynamics (Circuit
QED), which enabled strong coupling between qubits and microwave photons,
a critical feature for quantum information processing\cite{Blais_2004}.
Circuit QED has since played a crucial role not only in advancing
our understanding of light-matter interactions but also in the practical
development of quantum technologies. These include the realization
of single-qubit operations, the creation of two-qubit gates, and the
implementation of quantum error correction\cite{DiCarlo_2009}\cite{Ofek_2016}.
Moreover, Circuit QED has facilitated the exploration of hybrid quantum
systems, bridging the gap between different quantum platforms, such
as nitrogen-vacancy(NV) centers and semiconductor quantum dots, and
superconducting circuits\cite{Clerk_2020}\cite{Xiang_2013}. Today,
Circuit QED stands as a foundational architecture for quantum computing,
supporting the execution of quantum algorithms and the demonstration
of quantum supremacy\cite{Arute_2019}. This progress has brought
renewed attention to quantum transmission lines, which are poised
to play a vital role in the coherent transmission of quantum information.
As the complexity of quantum computing devices continues to increase,
new strategies for managing and controlling these intricate systems
are essential. A quantum approach to transmission lines may offer
the key insights needed to address these emerging challenges\cite{Parra_Rodriguez_2018,Parra_Rodriguez_2022,https://doi.org/10.48550/arxiv.2401.09120}.
For a detailed explanation, see.\cite{Vool_2017,Gu_2017,Blais_2021}.

In particle and condensed matter theory, a technique known as tensor
networks has received a great deal of attention in recent years. This
is a way of describing and treating entanglement as a kind of network
of tensors, closely related to renormalisation groups\cite{Okunishi_2022}. 

Continuous Multi-scale Entanglement Renormalisation Ansatz(cMERA)\cite{Haegeman_2013}
is a type of MERA, and MERA is a type of tensor networkwas is originally
a coarse-grained approach for determining the ground state of the
wavefunction of a quantum many-body system\cite{Vidal_2008}. The
relationship between its later coarse-graining direction and AdS space
was pointed out, and its correspondence with the holographic renormalisation
group in AdS/CFT in string theory was later discussed\cite{Swingle_2012}\footnote{However, this correspondence between MERA and AdS/CFT is a prediction
and does not currently show a clear agreement. This paper will assume
this prediction as correct. See \cite{Bao_2015,https://doi.org/10.48550/arxiv.1812.00529,https://doi.org/10.48550/arxiv.1807.02501,Erdmenger_2023}for
discussion after Swingle, especially \cite{Bao_2015}, which specifically
discusses the correspondence between MERA and AdS space.}.

Continuous MERA deals analytically with the discrete coarse-grained
steps of MERA, whose states can be obtained in an analytical form
using quantum-informative techniques to obtain a holographic space-time
picture\cite{Nozaki_2012}.

In order to explore the key indicators that can be seen by treating
complex quantum networks through holographic space-time, in this paper
we first discuss their holography through cMera in the quantisation
of the simplest transmission lines\footnote{Although not dealt with in this paper, a method called random tensor
network has recently been proposed and research on its holography
has progressed. See below for details\cite{Hayden_2016}.}.

This paper is organized as follows. In Section 2, we provide a detailed
discussion on the quantization of circuits, starting with the LCR
circuit and extending the theory to one-dimensional transmission lines.
We present the Hamiltonian and Lagrangian formulations and elaborate
on the quantization process. Section 3 explores the gravity-theoretic
description of a single transmission line, discussing the connection
between the MERA and AdS/CFT correspondence, and demonstrating the
application of continuous MERA (cMERA) to the transmission line theory.
In Section 4, we delve into the scenario where transmission lines
are coupled at $x=0$. We introduce the Lagrangian for interacting
transmission lines and derive the resulting equations of motion, discussing
the concept of scattering matrices for quantum networks. Section 5
examines holographic quantum circuit networks. Using the cMERA framework,
we perform the entanglement renormalisation group of transmission
lines with endpoints and obtain the corresponding holographic spacetime.
We analyze the relationship between the metric and the capacities
of the transmission lines and endpoints. Finally, in Section 6, we
present a discussion of the implications of our findings and potential
future research directions, considering how the results might be extended
to more complex quantum circuits and their holographic descriptions.
In addition, Apeendix A reviews the correspondence between MERA and
AdS space. Appendix B summarises the dimensional analysis for transmission
lines.

\section{Quantisation of circuits}

\begin{figure}
\includegraphics[width=1\textwidth,height=1\textheight,keepaspectratio]{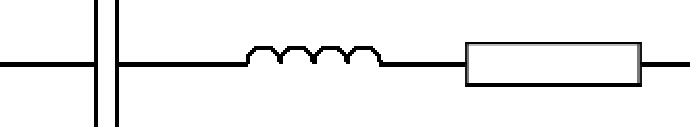}

\caption{LCR circuit\label{fig:LCR-circuit}}
\end{figure}

The quantum theory of transmission lines was first discussed by Yurk
and Denker\cite{Yurke_1984}. This section reviews quantisation of
LC circuits and transmission line.

If the voltage of the supply is taken as $V=0$ , the behaviour of
the LCR circuit is described by

\begin{equation}
L\ddot{Q}+R\dot{Q}+\frac{1}{C}Q=0,
\end{equation}
where $Q$ is electric charge, $L$ is inductance, $C$ is capacitance
and $R$ is resistance. If this is regarded as the equation of motion,
the resistance is the friction term (see Figure \ref{fig:LCR-circuit}).

Now consider the case of no resistance and the Lagrangian that gives
this equation is

\begin{equation}
\mathcal{L}=\frac{1}{2}L\dot{Q}^{2}-\frac{1}{2C}Q^{2}.
\end{equation}

The momentum is

\begin{equation}
\Phi=L\dot{Q},
\end{equation}
corresponding to the magnetic flux.

The Hamiltonian is

\begin{equation}
H=\frac{1}{2L}\Phi^{2}+\frac{1}{2C}Q^{2}.
\end{equation}

\begin{figure}
\subfloat[lossless transmission line\label{fig:lossless-transmission-line}]{\includegraphics[width=1\textwidth,height=1\textheight,keepaspectratio]{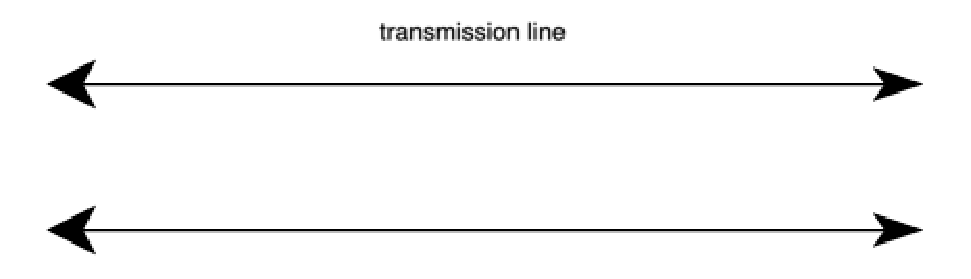}

}

\subfloat[Equivalent circuit of a lossless transmission line]{\includegraphics[width=1\textwidth,height=1\textheight,keepaspectratio]{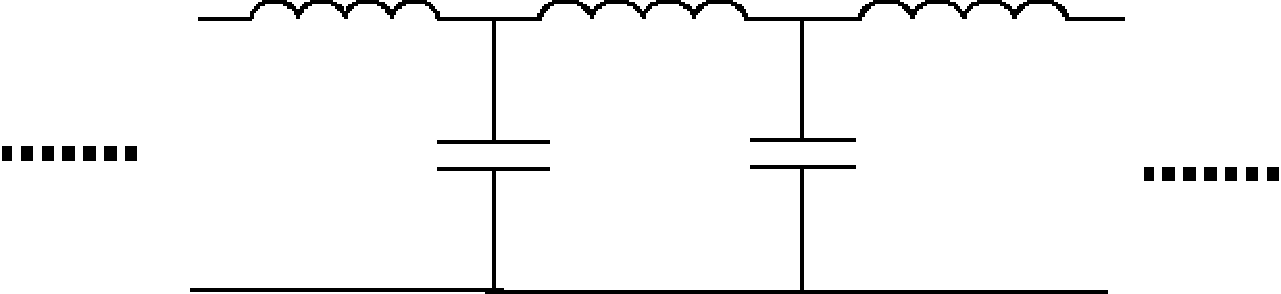}

}\caption{\label{fig:Equivalent-circuit-of}}
\end{figure}

Extending this to one-dimensional space is the theory of transmission
lines. A transmission line is a pair of wires, as shown in Figure
\ref{fig:lossless-transmission-line}, that transmit signals and power.
The equivalent circuit of a lossless transmission line is shown in
Figure \ref{fig:Equivalent-circuit-of}. We introduce inductance $L_{T}$
and capacitance $C_{T}$ per unit length. See Appendix B for these
dimensional analyses.

The total electric charge $Q(x,t)$ is then

\begin{equation}
Q(x,t)=\int_{-\infty}^{x}dx'q(x',t),
\end{equation}
using the charge density $q(x,t)$, and the electric current and voltage
are

\begin{equation}
I(x,t)=\frac{\partial Q(x,t)}{\partial t},
\end{equation}

\begin{equation}
V(x,t)=\frac{q(x,t)}{C_{T}}=\frac{\partial Q(x,t)}{C_{T}\partial x}.
\end{equation}

From the equivalent circuit of a lossless transmission line, we obtain
transmission line equation;

\begin{equation}
-\frac{\partial V(x,t)}{\partial x}=L_{T}\frac{\partial I}{\partial t},
\end{equation}

\begin{equation}
-\frac{\partial I(x,t)}{\partial x}=C_{T}\frac{\partial V(x,t)}{\partial t}.
\end{equation}

Then, we obtain

\begin{equation}
\frac{\partial^{2}Q}{C_{T}\partial x^{2}}=L_{T}\frac{\partial^{2}Q}{\partial t^{2}}.
\end{equation}

From this, the Hamiltonian density of the transmission line is

\begin{equation}
\begin{aligned}\mathcal{H} & =\frac{1}{2L_{T}}\Phi^{2}+\frac{1}{2C_{T}}q^{2}\\
 & =\frac{1}{2L_{T}}\Phi^{2}+\frac{1}{2C_{T}}\left(\frac{\partial Q}{\partial x}\right)^{2}.
\end{aligned}
\end{equation}

Using 
\begin{equation}
\Phi=L_{T}\frac{\partial Q}{\partial t},
\end{equation}
the Lagrangian density of the transmission line is

\begin{equation}
\mathcal{L}=\frac{1}{2}L_{T}\left(\frac{\partial Q}{\partial t}\right)^{2}-\frac{1}{2C_{T}}\left(\frac{\partial Q}{\partial x}\right)^{2}=\frac{1}{L_{T}}\left(\frac{1}{2}\Phi^{2}-\frac{1}{2}Z_{T}^{2}\left(\frac{\partial Q}{\partial x}\right)^{2}\right),
\end{equation}

\begin{equation}
Z_{T}^{2}\equiv\frac{L_{T}}{C_{T}}.
\end{equation}

The equation of motion is

\begin{equation}
L_{T}\frac{\partial^{2}Q(x,t)}{\partial t^{2}}-\frac{1}{C_{T}}\frac{\partial^{2}Q(x,t)}{\partial x^{2}}=0,
\end{equation}
which has the same form as the massless Klein-Gordon equation.

The speed at which the signal is transmitted is

\begin{equation}
v=\frac{1}{\sqrt{L_{T}C_{T}}}=\frac{1}{L_{T}}Z_{T}.
\end{equation}

As the transmission line can be described by the Klein-Gordon equation,
its quantisation can be directly applied to the usual quantum theory
of fields.

Since the canonical momentum corresponding to $Q$ is $\Phi$, the
canonical commutation relation is

\begin{equation}
[Q(x,t),\Phi(x',t)]=i\hbar\delta(x-x').
\end{equation}

This quantisation is not unnatural and can be explained from the quantisation
of the electromagnetic field. First, from Faraday's law of electromagnetic
induction, we obtain

\begin{equation}
\Phi(x,t)=\int_{-\infty}^{t}V(x,t')dt'=\int_{-\infty}^{t}\int^{x}E_{i}(x',t')dx'^{i}dt',
\end{equation}
where $E_{i}(x,t)$ is electric field.

Next, according to Ampere's law,

\begin{equation}
Q(x,t)=\int_{-\infty}^{t}I(x,t')dt'=\frac{1}{\mu_{0}}\int_{-\infty}^{t}\oint_{x}B_{i}(x',t')dx'^{i}dt,
\end{equation}
where $B_{i}(x,t)$ is magnetic flux density and $\mu_{0}$ is magnetic
permeability.

From $H_{i}(x,t)=\frac{1}{\mu_{0}}B_{i}(x,t),$ we obtain 
\begin{equation}
\begin{aligned}[][Q(x,t),\Phi(x',t)] & =\int_{-\infty}^{t}dt'\int_{-\infty}^{t}dt''\int^{x}dx''^{i}\oint_{x'}dx'''^{j}[H_{i}(x'',t'),E_{j}(x''',t'')]\\
 & =\int_{-\infty}^{t}dt'\int_{-\infty}^{t}dt''\int^{x}dx''^{i}\oint_{x'}dx'''^{j}i\hbar\delta^{3}(x''-x''')\delta(t'-t'')\\
 & =i\hbar\delta(x-x').
\end{aligned}
\end{equation}

It is also expanded by the creation and annihilation operator, which
produces a wavenumber representation of the physical variable as well
as the usual field quantisation;

\begin{equation}
Q(t,k)=\sqrt{\frac{\hbar}{2L\omega_{k}}}\left(a_{k}e^{i\left(kx+\omega_{k}t\right)}+a_{-k}^{\dagger}e^{-i\left(kx+\omega_{k}t\right)}\right),
\end{equation}

\begin{equation}
\Phi(t,k)=i\sqrt{\frac{\hbar L\omega_{k}}{2}}\left(a_{k}e^{i\left(kx+\omega_{k}t\right)}-a_{-k}^{\dagger}e^{-i\left(kx+\omega_{k}t\right)}\right),
\end{equation}

\begin{equation}
\omega_{k}=\frac{1}{\sqrt{L_{T}C_{T}}}k=\frac{1}{L_{T}}Z_{T}k,
\end{equation}

\begin{equation}
[a_{k},a_{k'}^{\dagger}]=\delta(k-k'),
\end{equation}

\begin{equation}
[Q(t,k),\Phi(t,k')]=i\hbar\delta(k+k').
\end{equation}

The Hamiltonian is

\begin{equation}
H=\frac{1}{L_{T}}\int dx\left(\frac{1}{2}\Phi(k)\Phi(-k)+\frac{1}{2}k^{2}Z_{T}^{2}Q(k)Q(-k)\right).
\end{equation}

\section{Gravity-theoretic description of a single transmission line}

Tensor networks are powerful tools for the efficient representation
of wavefunctions in quantum many-body systems. Among these, the Multi-scale
Entanglement Renormalisation Ansatz (MERA)\cite{Vidal_2008} has been
developed specifically to capture the scaling nature of critical states.
MERA effectively reconstructs the state of the system by resolving
entanglements at different scales. Continuous MERA (cMERA)\cite{Haegeman_2013}
extends this to continuous systems and is particularly important in
field theory and in research on critical phenomena.

As seen in the previous section, transmission lines can be described
by the one-dimensional Klein-Gordon equation, so the corresponding
AdS space-time description can be obtained by applying the continuous
MERA argument. In this section, we discuss this argument in transmission
line terms for the sake of review.

\subsection{cMERA}

cMERA applies scaling transformations and entanglement reduction operations
to the state to coarse-grain it while resolving entanglement at different
scales. The resulting state evolves from the IR state towards increased
entanglement as the scale increases. This scale direction corresponds
to the radial direction in AdS spacetime. This means that the state
of the transmission line is examined from a coarse-grained perspective,
from which the properties, for example, relation between Q-values
and variance of charge, of the transmission line can be extracted.

The direction in which this operation is performed is $u$, and its
evolution operator is the unitary operator as follows,

\begin{equation}
U(u_{1},u_{2})=P\exp\left[-i\int_{u_{1}}^{u_{2}}\left(K(u)+\mathbb{S}\right)du\right],
\end{equation}
where $P$ is the $u$-orderd product.

Firstly, $K(u)$ is the operator that entangles the state and is called
the entangler,

\begin{equation}
K(u)=\frac{1}{2\hbar}\int dkg(k,u)\left(Q(k)\Phi(-k)+\Phi(k)Q(-k)\right),
\end{equation}
where $g(k,u)$ is a scale-dependent variational parameter determined
to minimise the expectation value of energy, as discussed later. In
addition, $K$ is assumed to have a cut-off such that the mode does
not work except for those satisfying

\begin{equation}
k\leq\Lambda e^{u}.
\end{equation}

Next, $\mathbb{S}$ is the operator of scale transformation,

\begin{equation}
\mathbb{S}=-\frac{1}{2\hbar}\int dx[\Phi(x),xQ(x)+\frac{1}{2}\frac{\partial Q(x)}{\partial x}]_{+},
\end{equation}
where $[A,B]_{+}\equiv AB+BA$ and the action of $\mathbb{S}$ causes
a scale transformation as follows;

\begin{equation}
x\to xe^{u},\ Q(x)\to e^{-\frac{1}{2}u}Q(xe^{u}),\ \Phi(x)\to e^{-\frac{1}{2}u}\Phi(xe^{u}),
\end{equation}

\begin{equation}
k\to ke^{-u},\ Q(k)\to e^{\frac{1}{2}u}Q(ke^{-u}),\ \Phi(k)\to e^{-\frac{1}{2}u}\Phi(ke^{-u}).
\end{equation}

The action of $U$ transforms the physical variable as follows;

\begin{equation}
k\to ke^{-u},\ Q(k)\to e^{-f(k,u)}e^{\frac{1}{2}u}Q(ke^{-u}),\ \Phi(k)\to e^{f(k,u)}e^{-\frac{1}{2}u}\Phi(ke^{-u}),
\end{equation}

\begin{equation}
f(k,u)\equiv\int_{0}^{u}du'g(ke^{-u'},u').
\end{equation}

The state can be described as follows;

\begin{equation}
|\Psi(u)\rangle=U(u,u_{\mathrm{IR}})|\Omega\rangle=P\exp\left[i\int_{u_{IR}}^{u}\left(K(u)+\mathbb{S}\right)du\right]|\Omega\rangle,
\end{equation}
where $|\Omega\rangle=|\Psi(u_{\mathrm{IR}})\rangle$ is the IR limit
state and the scale is $u_{\mathrm{IR}}$. We set $u_{\mathrm{IR}}=-\infty$.

Because $\mathrm{IR}$ limit is scale invariant, we obtain

\begin{equation}
\mathbb{S}|\Omega\rangle=0.
\end{equation}

Also, we set the scale at UV limit, $u_{\mathrm{UV}}=0$, $|\Psi\rangle\equiv|\Psi(u_{\mathrm{UV}})\rangle=|\Psi(0)\rangle.$

Similarly, we obtain in UV limit,

\begin{equation}
|\Psi\rangle=U(0,u)|\Psi(0)\rangle=P\exp\left[i\int_{u}^{0}\left(K(u)+\mathbb{S}\right)du\right]|\Psi(u)\rangle.
\end{equation}

In the IR limit, we want to take a state $|\Omega\rangle$ with zero
entanglement entropy as the state before taking entanglement.

We shall therefore take the following separable state: 
\begin{equation}
\langle Q(x)|\Omega\rangle\propto\prod_{x}e^{\frac{i}{2}L_{T}\omega_{\Lambda}Q(x)^{2}}.
\end{equation}
where $\omega_{\Lambda}$ is $\omega_{\Lambda}=\frac{\Lambda}{L_{T}}Z_{T}$
and $\Lambda$ is a UV cutoff parameter.

Such a state is made up of a direct product of local operators, so
the Neumann entropy is zero.

This state satisfies

\begin{equation}
\left(\sqrt{L_{T}\omega_{\Lambda}}Q(x)+\frac{i}{\sqrt{L_{T}\omega_{\Lambda}}}\Phi(x)\right)|\Omega\rangle=0,
\end{equation}
as the condition that the state is annihilated by an annihilation
operator consisting of local operators.

From this definition of $\Omega$, the following equation is obtained:

\begin{equation}
\langle\Omega|Q(k)Q(k')|\Omega\rangle=\frac{\hbar}{2L_{T}\omega_{\Lambda}}\delta(k+k'),
\end{equation}

\begin{equation}
\langle\Omega|\Phi(k)\Phi(k')|\Omega\rangle=\frac{L_{T}\hbar\omega_{\Lambda}}{2}\delta(k+k').
\end{equation}

In the following, we shall discuss this using an interaction representation:

\begin{equation}
K_{I}(u)\equiv e^{iu\mathbb{S}}K(u)e^{-iu\mathbb{S}}=\frac{1}{2\hbar}\int dkg(ke^{-u},u)\left(Q(k)\Phi(-k)+\Phi(k)Q(-k)\right),
\end{equation}

\begin{equation}
H(u_{IR})=e^{iu\mathbb{S}}He^{-iu\mathbb{S}},
\end{equation}

\begin{equation}
U(u_{1},u_{2})=e^{-iu_{1}\mathbb{S}}P\exp\left[-i\int_{u_{2}}^{u_{1}}K_{I}(s)ds\right]e^{iu_{2}\mathbb{S}},
\end{equation}

\begin{equation}
|\Psi_{I}(u)\rangle\equiv e^{iu\mathbb{S}}|\Psi(u)\rangle=P\exp\left[-i\int_{u_{IR}}^{u}K_{I}(s)ds\right]|\Omega\rangle.
\end{equation}

We now determine $g(k,u)$ such that the energy expectation

\begin{equation}
E=\langle\psi|H|\psi\rangle=\langle\Omega|H(u_{IR})|\Omega\rangle
\end{equation}
is minimised from the variational method.

Let us assume that $g(k,u)$ is like

\begin{equation}
g(k,u)=\chi(u)\Theta\left(1-\left|\frac{k}{\Lambda}\right|\right).
\end{equation}

Here, for simplicity, it is assumed that the mode is independent.

From this assumption, the energy expectation value is

\begin{equation}
E=\int dk\frac{1}{4}\left(e^{2f(k,u_{IR})}\omega_{\Lambda}+\frac{\omega_{k}^{2}}{\omega_{\Lambda}}e^{-2f(k,u_{IR})}\right).
\end{equation}

Therefore, differentiating by $\chi$ gives

\begin{equation}
\frac{\delta E}{\delta\chi(u)}=\int_{|k|\leq\Lambda e^{u}}dk\frac{1}{2}\left(e^{2f(k,u_{IR})}\omega_{\Lambda}-\frac{\omega_{k}^{2}}{\omega_{\Lambda}}e^{-2f(k,u_{IR})}\right).
\end{equation}

Therefore,

\begin{equation}
f(k,u_{\mathrm{IR}})=\frac{1}{2}\log\frac{\omega_{k}}{\omega_{\Lambda}}
\end{equation}
is obtained as a condition for this to be $\text{\ensuremath{\frac{\delta E}{\delta g(u)}}=0}$,
and consequently $f(k,u)$ and $\chi(u)$ is obtained as

\begin{equation}
f(k,u)=\frac{1}{2}u,(|k|<\ensuremath{\Lambda e^{u})},
\end{equation}

\begin{equation}
\chi(u)=\frac{1}{2}.
\end{equation}

Using this, the state at some $u$ is

\begin{equation}
|\Psi(u)\rangle=e^{-iu\mathbb{S}}P\exp\left[\frac{1}{2\hbar}\int^{-\log\Lambda/|ke^{-u}|}dk\frac{1}{2}\left(Q(k)\Phi(-k)+\Phi(k)Q(-k)\right)\right]|\Omega\rangle,
\end{equation}

\begin{equation}
\frac{\partial}{\partial u}|\Psi(u)\rangle=-i\left(\mathbb{S}+K_{I}\right)|\Psi(u)\rangle.\label{eq:uPsi}
\end{equation}

\subsection{$\mathrm{AdS/CFT}$}

The correspondence between MERA and AdS/CFT has already been discussed\cite{Nozaki_2012}.
There it is shown that the parameter $u$-direction of the cMera entanglement
renormalisation corresponding to the layer of the MERA index corresponds
to the radial direction of the AdS/CFT:

\begin{equation}
ds_{\mathrm{AdS}_{3}}^{2}=\frac{dz^{2}+dx^{2}-dt^{2}}{z^{2}}=du^{2}+\frac{e^{-2u}}{\epsilon^{2}}\left(dx^{2}-dt^{2}\right)
\end{equation}
where $(z,x,t$) are coordinates for Poincare $\mathrm{AdS_{3}}$
metric. We define $z=\epsilon e^{u}$. Also in this paper we take
$(++-)$ as the sign of the metric.

The state we obtained contained a parameter $u$ of the entanglement
renormalisation group. cMera and the AdS/CFT correspondence, it is
necessary to at least check whether the metric in the $u$ direction
reproduces the $g_{uu}$ of the AdS metric\footnote{The relations between Mera and the AdS space are reviewed in Appendix
A.}.

Therefore, we extract metric from quantum states using the information
geometry approach.

First, we introduce

\begin{equation}
D_{Q}(\theta)=1-|\langle\psi(\theta)|\psi(\theta+d\theta)\rangle|^{2},
\end{equation}
which we call the quantum distance (Bars distance, Fubini-Study distance).

The metric is introduced by expanding this distance $D_{Q}(\theta)$
up to the second order of $d\theta$:

\begin{equation}
D_{Q}(\theta)\approx\frac{1}{2}\left(\chi_{\mu\nu}+\chi_{\nu\mu}\right)d\theta^{\mu}d\theta^{\nu}\equiv g_{\mu\nu}d\theta^{\mu}d\theta^{\nu}
\end{equation}

\begin{equation}
\chi_{\mu\nu}=\langle\partial_{\mu}\psi(\theta)|\partial_{\nu}\psi(\theta)\rangle-\langle\partial_{\mu}\psi(\theta)|\psi(\theta)\rangle\langle\psi(\theta)|\partial_{\nu}\psi(\theta)\rangle.
\end{equation}

Using (\ref{eq:uPsi}), we obtain 
\begin{equation}
\begin{aligned}g_{uu}(u) & =\langle\Psi_{I}(u)|K_{I}(u)^{2}|\Psi_{I}(u)\rangle-\langle\Psi_{I}(u)|K_{I}(u)|\Psi_{I}(u)\rangle^{2}\\
 & =\chi(u)^{2}=\frac{1}{4}
\end{aligned}
\end{equation}
and is consistent with the result that $g_{uu}$ is a constant.

It has been stated above that the quantisation of a transmission line
can be placed directly in the context of $\mathrm{AdS_{3}}$/$\mathrm{CFT}_{2}$
because it is a scalar field itself. What this argument does, namely,
is to argue that when the direction of coarse-graining is considered,
an $\mathrm{AdS}$ space appears in which the direction of coarse-graining
is added to the information space. It is not clear how to generalise
this further to actions where transmission lines interact with $\mathrm{LC}$
circuits by mutual inductance etc., but in the situation where one
transmission line interacts with an $\mathrm{LC}$ circuit at $x=0$,
the transmission line can be regarded as a resistance at $x=0$ and
the whole combination of transmission lines and LC circuits can be
represented as an LCR circuit. In the quantum theory of the $\mathrm{LCR}$
circuit, the behaviour of the charge dispersion can be classified
according to $Q$-values. It is interesting to see how this classification
by $Q$-value is described on the $\mathrm{AdS}$ side. Therefore,
this case is explained in the next section.

\section{When transmission lines are coupled at $x=0$}

The discussion so far has been about free scalar fields, which is
almost self-evident in the context of cMERA. In order to develop a
quantum network theory, it is necessary to discuss models that have
not been dealt with in the discussion of cMera up to now. As a first
step, we would like to consider the case where several parallel transmission
lines are coupled at $x=0$. Such systems are the basis of complex
quantum networks.

Suppose there are several parallel transmission lines labelled with
$i$, which are joined at $x=0$. The Lagrangian density is

\begin{equation}
\mathcal{L}=\delta(x)\tilde{L}+\theta(x)\sum_{i}\mathcal{L}_{i}
\end{equation}
where $\mathcal{L}_{i}$ represents the $i$-th transmission line;
\begin{equation}
\mathcal{L}_{i}=\frac{1}{2}L_{i}\left(\frac{\partial Q_{i}}{\partial t}\right)^{2}-\frac{1}{2C_{i}}\left(\frac{\partial Q_{i}}{\partial x}\right)^{2}
\end{equation}
and $\tilde{L}$ represents the term with which the transmission lines
interact at $x=0$\footnote{The difficulties of using the Dirac function for endpoints and avoiding
this are discussed in \cite{Parra_Rodriguez_2022,Forestiere_2024}.}, using mutual inductance $L_{ij}$ and capacitance $C_{ij}$,

\begin{equation}
\tilde{L}=\frac{1}{2}\left(\sum_{ij}L_{ij}\frac{dQ_{i}}{dt}\frac{dQ_{j}}{dt}-\frac{1}{2C_{ij}}Q_{i}Q_{j}\right).
\end{equation}

The equation of motion for this Lagrangian is

\begin{equation}
\delta(x)\sum_{j}\left[L_{ij}\frac{\partial^{2}Q_{j}}{\partial t^{2}}+\frac{\partial Q_{j}}{C_{ij}}\right]+\theta(x)L_{i}\frac{\partial^{2}Q_{i}}{\partial t^{2}}-\frac{1}{C_{i}}\frac{\partial}{\partial x}\left(\theta(x)\frac{\partial Q_{i}}{\partial x}\right)=0.
\end{equation}

In $x>0$, this equations are Klein-Gordon equations

\begin{equation}
L_{i}\frac{\partial^{2}Q_{i}}{\partial t^{2}}-\frac{1}{C_{i}}\frac{\partial^{2}Q_{i}}{\partial x^{2}}=0.
\end{equation}

At the endpoints, by integrating over $-\epsilon\leq x\leq\epsilon$
and then take $\epsilon\to0$, we obtain

\begin{equation}
\sum_{j}\left[L_{ij}\frac{d^{2}Q_{j}}{dt^{2}}+\frac{Q_{j}}{C_{ij}}\right]-\frac{1}{C_{i}}\left.\frac{\partial Q_{i}}{\partial x}\right|_{x=0+}=0.
\end{equation}

Since $Q_{i}$ satisfies the Klein-Gordon equation at $x\geq0$, the
general solutions $Q_{\mathrm{in}}$, $Q_{\mathrm{out}}$ represent
the outgoing and incoming solutions from $x=0$;

\begin{equation}
Q_{i}(x,t)=Q_{i}^{\mathrm{in}}\left(\frac{x}{v_{i}}+t\right)+Q_{i}^{\mathrm{out}}\left(-\frac{x}{v_{i}}+t\right),
\end{equation}

\begin{equation}
v_{i}\equiv\frac{1}{\sqrt{L_{i}C_{i}}}.
\end{equation}

By partial differentiation of this equation with $x,t$, we obtain

\begin{equation}
-\frac{1}{C_{i}}\left.\frac{\partial Q}{\partial x}\right|_{x=0+}=R_{i}\left(\frac{d}{dt}Q_{i}-2\frac{d}{dt}Q_{i}^{\mathrm{in}}\right),
\end{equation}

\begin{equation}
R_{i}\equiv\frac{1}{C_{i}v_{i}}=\sqrt{\frac{L_{i}}{C_{i}}}.
\end{equation}

Therefore, the equation of motion at the endpoint becomes

\begin{equation}
\sum_{j}\left[L_{ij}\frac{d^{2}Q_{j}}{dt^{2}}+\frac{Q_{j}}{C_{ij}}\right]+R_{i}\frac{d}{dt}Q_{i}=2R_{i}\frac{dQ^{\mathrm{in}}}{dt}\label{eq:Mutal Heisenberg}
\end{equation}
where $R_{i}\frac{d}{dt}Q_{i}$ represents Ohm's law at $x=0$ and
$2R_{i}\frac{dQ^{\mathrm{in}}}{dt}$ on the right-hand side represents
the noise voltage transmitted from the transmission line. The whole
equation can therefore be regarded as the Langevin equation.

By mode expansion, we obtain

\begin{equation}
Q_{i}^{\mathrm{in}}(t)=\int dk\sqrt{\frac{\hbar}{2L_{i}\omega_{k}}}\left(a_{k}^{\mathrm{in}}e^{i\omega_{k}t}+a_{-k}^{\mathrm{in}\dagger}e^{-i\omega_{k}t}\right),
\end{equation}

\begin{equation}
\Phi_{i}^{\mathrm{in}}(t)=\Phi_{i}(0,t)=\int dk\sqrt{\frac{L_{i}\hbar\omega_{k}}{2}}\left(a_{k}^{\mathrm{in}}e^{i\omega_{k}t}-a_{-k}^{\mathrm{in}\dagger}e^{-i\omega_{k}t}\right),
\end{equation}

\begin{equation}
\omega_{k}=vk,
\end{equation}
where $\Phi_{i}$ is counjugate momentum of $Q_{i}$ as in Section
2.

From (\ref{eq:Mutal Heisenberg}), we obtain $Q_{i}$ in terms of
$Q_{i}^{\mathrm{in}}(t).$

Also, $Q^{\mathrm{out}}$(t) is solved from 
\begin{equation}
Q_{i}^{\mathrm{out}}(t)=Q_{i}(0,t)-Q_{i}^{\mathrm{in}}(t).
\end{equation}

Then, $a_{k}^{\mathrm{out}}$ can be described by $a_{k}^{\mathrm{in}}$.

For linear systems, this can be written as a linear combination,

\begin{equation}
a_{i,k}^{\mathrm{out}}=\sum_{j}S_{ij}a_{j,k}^{\mathrm{in}}
\end{equation}

$S_{ij}$ can be regarded as a quantum network version of the scattering
matrix.

Now, for simplicity's sake, let's assume one transmission line:

\begin{equation}
\mathcal{L}=\delta(x)\left[\frac{1}{2}L\dot{Q}^{2}-\frac{1}{2C}Q^{2}\right]+\theta(x)\left[\frac{1}{2}L_{T}\left(\frac{\partial Q}{\partial t}\right)^{2}-\frac{1}{2C_{T}}\left(\frac{\partial Q}{\partial x}\right)^{2}\right].
\end{equation}

\begin{figure}
\includegraphics[width=1\textwidth,height=1\textheight,keepaspectratio]{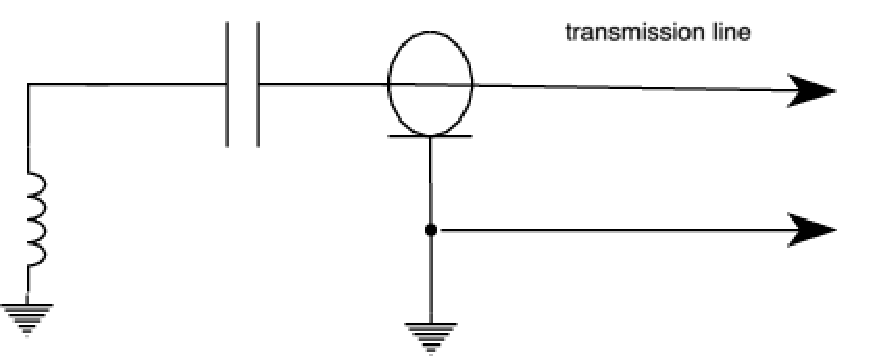}

\caption{Transmission line with endpoint\label{fig:Transmission-line-with}}
\end{figure}

As can be seen from Figure. \ref{fig:Transmission-line-with} which
shows a diagram of the corresponding transmission line, this is a
transmission line directly connected to the LC circuit at its end.
This transmission line acts as a low impedance source. In such systems,
the transmission line is responsible for transmitting the signals
from the LC circuit at the end of the line. Efficient signal transmission
in this setup requires impedance matching between the LC circuit and
the transmission line, which is often discussed in terms of Q-values.

The equation of motion at the endpoint is

\begin{equation}
L\frac{d^{2}}{dt^{2}}Q+R\frac{dQ}{dt}+\frac{Q}{C}=2R\frac{dQ^{\mathrm{in}}}{dt},
\end{equation}

\begin{equation}
R=\sqrt{\frac{L_{T}}{C_{T}}}.
\end{equation}

Using mode expansion 
\begin{equation}
2R\frac{dQ^{\mathrm{in}}}{dt}=\int dk\sqrt{\frac{\hbar\omega_{k}}{2C_{T}}}\left(a_{k}^{\mathrm{in}}e^{i\omega_{k}t}-a_{k}^{\mathrm{in}\dagger}e^{-i\omega_{k}t}\right)
\end{equation}
we obtain 
\begin{equation}
L\ddot{Q}+R\dot{Q}+\frac{Q}{C}=\int dk\sqrt{\frac{\hbar\omega_{k}}{2C_{T}}}\left(a_{k}^{\mathrm{in}}e^{i\omega_{k}t}-a_{k}^{\mathrm{in}\dagger}e^{-i\omega_{k}t}\right).
\end{equation}

From this, we obtain 
\begin{equation}
Q(t)=\int dk\sqrt{\frac{\hbar\omega_{k}}{2C_{T}}}\left(\frac{-1}{\omega_{k}^{2}L-1/C-i\omega_{k}R}a_{k}^{\mathrm{in}}e^{+i\omega_{k}t}-h.c.\right).\label{eq: Q(t)with endpoint}
\end{equation}

From this result, $a^{\mathrm{out}}$ is obtained as

\begin{equation}
a^{\mathrm{out}}(\omega)=\frac{\omega^{2}L-1/C+i\omega R}{\omega^{2}L-1/C-i\omega R}a^{\mathrm{in}}(\omega).
\end{equation}

From this, the scattering matrix is

\begin{equation}
S=\frac{\omega^{2}L-1/C+i\omega R}{\omega^{2}L-1/C-i\omega R}.
\end{equation}

Let's quantize the system using the procedure of canonical quantization
and define the vacuum to satisfy condition

\begin{equation}
a^{\mathrm{in}}(\omega)|0\rangle=0.
\end{equation}

From this, the expectation value and variance of $Q(t)$ are 
\begin{equation}
\langle0|Q(t)|0\rangle=0,
\end{equation}

\begin{equation}
\Delta Q^{2}=\langle0|Q^{2}(t)|0\rangle=\frac{R}{2}\int_{0}^{\infty}\frac{\hbar\omega_{k}}{L^{2}\left(\omega_{k}^{2}-\omega_{0}^{2}\right)^{2}+\omega_{k}^{2}R^{2}}d\omega_{k}\label{eq:variance}
\end{equation}
where 
\begin{equation}
\omega_{0}\equiv\frac{1}{\sqrt{LC}}.
\end{equation}

For a better understanding of the operation and properties of quantum
circuits, it is important to analyse the behaviour of charges precisely.
Therefore, Q-values $\mathbb{Q}$ are introduced as a new parameter
to classify the dispersion of charges in transmission lines in detail.
This will clarify the physical description on the AdS side and enable
a clearer understanding of the characteristics of quantum circuits:

\begin{equation}
\mathbb{Q}\equiv\frac{L}{R}\omega_{0}=\sqrt{\frac{L}{C}}\sqrt{\frac{C_{T}}{L_{T}}}.
\end{equation}
With the $\mathbb{Q}$, equation (\ref{eq:variance}) becomes 
\begin{equation}
\Delta Q^{2}=\frac{1}{2R}\int_{0}^{\infty}\frac{\hbar\mathbb{Q}_{k}}{\left(\mathbb{Q}_{k}^{2}-\mathbb{Q}^{2}\right)^{2}+\mathbb{Q}_{k}^{2}}d\mathbb{Q}_{k},
\end{equation}

\begin{equation}
\mathbb{Q}_{k}\equiv\frac{L}{R}\omega_{k}.
\end{equation}

Therefore, depending on the value of the Q-value, the behaviour of
the variance is as follows;

\begin{equation}
\Delta Q^{2}=\begin{cases}
\frac{\hbar}{2R}\frac{\pi+2\arctan\left(\frac{-1+2\mathbb{Q}^{2}}{\sqrt{-1+4\mathbb{Q}^{2}}}\right)}{2\sqrt{-1+4\mathbb{Q}^{2}}} & (\mathbb{Q}>\frac{1}{2})\\
\frac{\pi\hbar}{2R} & (\mathbb{Q}=\frac{1}{2})\\
\frac{\hbar}{2R}\frac{i\pi+2\mathrm{\arctan\left(\frac{1-2\mathbb{Q}^{2}}{\sqrt{1-4\mathbb{Q}^{2}}}\right)}}{2\sqrt{1-4\mathbb{Q}^{2}}} & (\mathbb{Q}<\frac{1}{2})
\end{cases}\label{eq:QVariance}
\end{equation}

In particular, in the $\mathbb{Q}\to\infty$ limit, using Cauchy's
integral theorem, the variance is

\begin{equation}
\Delta Q^{2}=\frac{\pi\hbar}{4R},
\end{equation}
which represents the minimum uncertainty state in the quantum mechanics
of harmonic oscillators.

\section{Holographic quantum circuit networks}

We discuss states in quantum networks with endpoints at $x=0$ using
the cMera introduced in the previous section and describe this in
holographic terms.

\subsection{cMERA}

The basic strategy is considered to be similar to the cMera argument
for scalar fields.

From the Lagrangian;

\begin{equation}
\mathcal{L}=\delta(x)\left[\frac{1}{2}L\dot{Q}^{2}-\frac{1}{2C}Q^{2}\right]+\theta(x)\left[\frac{1}{2}L_{T}\left(\frac{\partial Q}{\partial t}\right)^{2}-\frac{1}{2C_{T}}\left(\frac{\partial Q}{\partial x}\right)^{2}\right]
\end{equation}
we obtain the Hamiltonian 
\begin{equation}
\begin{aligned}H= & \int dkdk'\left(\frac{1}{2}\left(\frac{1}{L}+\frac{1}{L_{T}}\delta(k+k')\right)\Phi(k)\Phi(k')\right.\\
 & \left.+\frac{1}{2}\left(L\omega^{2}+L_{T}\omega_{k}^{2}\delta(k+k')\right)Q(k)Q(k')\right)
\end{aligned}
\end{equation}
where

\begin{equation}
\Phi\equiv L\dot{Q},
\end{equation}

\begin{equation}
\Phi_{T}=L_{T}\frac{\partial Q}{\partial t},
\end{equation}

\begin{equation}
\omega_{k}^{2}=\frac{k^{2}}{L_{T}C_{T}},
\end{equation}

\begin{equation}
\omega^{2}=\frac{1}{LC}.
\end{equation}

As in the previous section, we consider the expectation value of energy

\begin{equation}
E=\langle\Psi|H|\text{\ensuremath{\Psi}}\rangle
\end{equation}

\begin{equation}
|\Psi\rangle=P\exp\left[-i\int_{-u_{IR}}^{0}K_{I}(s)ds\right]|\Omega\rangle
\end{equation}
where,

\begin{equation}
\mathbb{S}=-\frac{1}{2}\int dx\left(x\Phi\frac{\partial Q}{\partial x}+x\frac{\partial Q}{\partial x}\Phi+\frac{1}{2}\left(Q\Phi+\Phi Q\right)\right),
\end{equation}

\begin{equation}
K=\frac{1}{2}\int dkg(k,u)\left(Q(k)\Phi(-k)+\Phi(k)Q(-k)\right),
\end{equation}
and find $g(k,u)$ such the energy $E$ this is minimised.

We assume $g(k,u)$ is 
\begin{equation}
g(k,u)=\chi(u)\Theta\left(1-\left|\frac{k}{\Lambda}\right|\right).
\end{equation}

From the Heisenberg formalism,

\begin{equation}
E=\langle\Psi|H|\Psi\rangle=\langle\Omega|H(u_{IR})|\Omega\rangle.
\end{equation}

Squeezing causes the Hamiltonian to undergo changes and individual
elements are squeezed as follows;

\begin{equation}
Q(k)\to e^{-f(k,u)}e^{\frac{1}{2}u}Q(ke^{-u}),
\end{equation}

\begin{equation}
\Phi(k)\to e^{f(k,u)}e^{-\frac{1}{2}u}\Phi(ke^{-u}),
\end{equation}

\begin{equation}
f(k,u)\equiv\int_{0}^{u}du'g(ke^{-u'},u').
\end{equation}

From the above,

\begin{equation}
\langle\Omega|Q(k)Q(k')|\Omega\rangle=\frac{1}{2L_{T}\omega_{\Lambda}}\delta(k+k'),
\end{equation}

\begin{equation}
\langle\Omega|\Psi(k)\Psi(k')|\Omega\rangle=\frac{L_{T}\omega_{\Lambda}}{2}\delta(k+k').
\end{equation}
Using this, the expection value of energy is

\begin{equation}
\begin{aligned}E= & \int dk\frac{1}{4}\left(\left(1+\frac{L_{T}\Lambda}{L}\right)e^{2f(k,u_{IR})}\omega_{\Lambda}+\left(\omega_{k}^{2}+\frac{L}{L_{T}}\omega^{2}\Lambda\right)e^{-2f(k,u_{IR})}\frac{1}{\omega_{\Lambda}}\right)\end{aligned}
\end{equation}
and differentiating by $\chi$

\begin{equation}
\frac{\delta E}{\delta\chi(u)}=\int_{|k|\leq\Lambda e^{u}}\frac{1}{2}\left(\left(1+\frac{L_{T}\Lambda}{L}\right)e^{2f(k,u_{IR})}\omega_{\Lambda}-\left(\omega_{k}^{2}+\frac{L}{L_{T}}\omega^{2}\Lambda\right)e^{-2f(k,u_{IR})}\frac{1}{\omega_{\Lambda}}\right),
\end{equation}
the minimum value for $\chi$ is

\begin{equation}
f(k,u_{\mathrm{IR}})=\frac{1}{2}\log\frac{\omega_{k}}{\omega_{\Lambda}}\sqrt{1+\frac{L}{L_{T}}\frac{\omega^{2}}{\omega_{k}^{2}}\Lambda}=\frac{1}{2}\log\frac{k}{\Lambda}\sqrt{1+\frac{L}{L_{T}}\frac{k^{2}}{\Lambda^{2}}\Lambda}.\label{eq:f(k,u)}
\end{equation}
Using

\begin{equation}
f(k,u_{IR})=\int_{0}^{u_{IR}}g(ke^{-s},s)ds=\int_{0}^{u_{IR}}\chi(s)\Theta\left(1-\left|\frac{ke^{-s}}{\Lambda}\right|\right)ds=\int_{0}^{-\log\Lambda/|k|}\chi(s)ds,
\end{equation}
we obtain

\begin{equation}
\chi(s)=\frac{1}{2\left(1+\frac{L_{T}}{L}\frac{1}{\Lambda e^{-2s}}\right)}.
\end{equation}

\subsection{Variance}

Let us now examine the variance of the states obtained by Mera:

\begin{equation}
\Delta Q(t)=\langle\Psi|Q(t)^{2}|\Psi\rangle-\langle\Psi|Q(t)|\Psi\rangle^{2}.
\end{equation}
we second term in this equation is obviously zero. From equation (\ref{eq: Q(t)with endpoint}),
we obtain 
\begin{equation}
\langle\Psi|Q(0,t)Q(0,t)|\Psi\text{\ensuremath{\rangle}}=\int_{0}^{\infty}\frac{\omega}{\omega^{2}+\frac{L^{2}}{R^{2}}(\omega^{2}-\omega_{0}^{2})}\langle\Psi|a_{k}^{in}a_{-k}^{in}|\Psi\rangle d\omega.
\end{equation}
Using 
\begin{equation}
\langle\Psi|a_{k}^{\mathrm{in}}a_{-k}^{\mathrm{in}}|\Psi\rangle=\langle\Omega|a_{k}^{\mathrm{in}}(u)a_{-k}^{\mathrm{in}}(u)|\Omega\rangle,
\end{equation}

\begin{equation}
a_{k}^{\mathrm{in}}(u)=U^{\dagger}a_{k}^{\mathrm{in}}U=a_{k}^{\mathrm{in}}e^{-f(ke^{u},u)}e^{\frac{1}{2}u},
\end{equation}
we obtain: 
\begin{equation}
\langle\Psi|Q(0,t)Q(0,t)|\Psi\text{\ensuremath{\rangle}}=\frac{1}{2R}\int_{0}^{\infty}\frac{\hbar\mathbb{Q}_{k}}{\left(\mathbb{Q}_{k}^{2}-\mathbb{Q}^{2}\right)^{2}+\mathbb{Q}_{k}^{2}}e^{-2f(k,u_{IR})}e^{u_{IR}}d\mathbb{Q}_{k}.
\end{equation}
From (\ref{eq:f(k,u)}), this equation is 
\begin{equation}
\langle\Psi|Q(0,t)Q(0,t)|\Psi\text{\ensuremath{\rangle}}=\frac{1}{2R}\int_{0}^{\infty}\frac{\mathbb{Q}_{k}}{\left(\mathbb{Q}_{k}^{2}-\mathbb{Q}^{2}\right)^{2}+\mathbb{Q}_{k}^{2}}\left(\frac{\omega_{\Lambda}}{\omega_{k}}\frac{1}{\sqrt{1+\frac{L}{L_{T}}\frac{\omega^{2}}{\omega_{k}^{2}}\Lambda}}\right)d\mathbb{Q}_{k},
\end{equation}

\begin{equation}
=\frac{1}{2R}\int_{0}^{\infty}\frac{\mathbb{Q}_{k}}{\left(\mathbb{Q}_{k}^{2}-\mathbb{Q}^{2}\right)^{2}+\mathbb{Q}_{k}^{2}}\left(\frac{\frac{\omega_{\Lambda}}{\omega}}{\sqrt{\frac{L}{L_{T}}\Lambda}\sqrt{1+\frac{1}{\Lambda}\frac{L_{T}}{L}\frac{\omega_{k}^{2}}{\omega^{2}}}}\right)d\mathbb{Q}_{k},
\end{equation}

\begin{equation}
=\frac{1}{2R}\int_{0}^{\infty}\frac{\mathbb{Q}_{k}}{\left(\mathbb{Q}_{k}^{2}-\mathbb{Q}^{2}\right)^{2}+\mathbb{Q}_{k}^{2}}\left(\frac{\frac{\mathbb{Q}_{\Lambda}}{\mathbb{Q}}}{\sqrt{\frac{L}{L_{T}}\Lambda}\sqrt{1+\frac{1}{\Lambda}\frac{L_{T}}{L}\frac{\mathbb{Q}_{k}^{2}}{\mathbb{Q}^{2}}}}\right)d\mathbb{Q}_{k},
\end{equation}

\begin{equation}
\end{equation}

where $\mathbb{Q}_{\Lambda}=\frac{L}{R}\omega_{\Lambda},\omega_{\Lambda}=\frac{\Lambda}{\sqrt{L_{T}C_{T}}}.$

Expaanding $\frac{1}{\Lambda}\frac{C_{T}}{C}\frac{\mathbb{Q}_{\Lambda}}{\mathbb{Q}_{k}}\ll1$
,

\begin{equation}
\frac{1}{2R}\frac{\mathbb{Q}_{k}}{\left(\mathbb{Q}_{k}^{2}-\mathbb{Q}^{2}\right)^{2}+\mathbb{Q}_{k}^{2}}\left(\frac{\frac{\mathbb{Q}_{\Lambda}}{\mathbb{Q}}}{\sqrt{\frac{L}{L_{T}}\Lambda}\sqrt{1+\frac{1}{\Lambda}\frac{L_{T}}{L}\frac{\mathbb{Q}_{k}^{2}}{\mathbb{Q}^{2}}}}\right)=\frac{1}{2R}\frac{\mathbb{Q}_{k}}{\left(\mathbb{Q}_{k}^{2}-\mathbb{Q}^{2}\right)^{2}+\mathbb{Q}_{k}^{2}}\frac{\frac{\mathbb{Q}_{\Lambda}}{\mathbb{Q}}}{\sqrt{\frac{L}{L_{T}}\Lambda}}\left(1-\frac{1}{2\Lambda}\frac{L_{T}}{L}\frac{\mathbb{Q}_{k}^{2}}{\mathbb{Q}^{2}}+\dots\right)
\end{equation}

In this case, with $\mathbb{Q}>1/2$, the variance is as follows.
\begin{equation}
\begin{aligned}\Delta Q^{2} & =\frac{\hbar}{2R}\gamma(\mathbb{Q},\Lambda)\left(\frac{\pi+2\mathrm{arctan}\left(\frac{-1+2\mathbb{Q}^{2}}{\sqrt{-1+4\mathbb{Q}^{2}}}\right)}{2\sqrt{-1+4\mathbb{Q}^{2}}}\right.\\
 & \left.-\frac{1}{2\Lambda}\frac{L_{T}}{L}h(\mathbb{Q})\right),\\
\gamma(\mathbb{Q},\Lambda) & \equiv\frac{\mathbb{Q}_{\Lambda}}{\mathbb{Q}}\sqrt{\frac{L_{T}}{L\Lambda}}=\frac{1}{\mathbb{Q}}\sqrt{\frac{L\Lambda}{L_{T}}},\\
h(\mathbb{Q}) & \equiv\int_{0}^{\infty}\frac{\mathbb{Q}_{k}^{3}}{(\mathbb{Q}_{k}^{2}-\mathbb{Q}^{2})^{2}+\mathbb{Q}_{k}^{2}}d\mathbb{Q}_{k}
\end{aligned}
\end{equation}

In limit $\Lambda\to\infty$ while keeping $\gamma(\mathbb{Q},\Lambda)=\gamma$

\begin{equation}
\Delta Q^{2}=\frac{\gamma\hbar}{2R\mathbb{}}\frac{\pi+2\mathrm{arctan}\left(\frac{-1+2\mathbb{Q}}{\sqrt{-1+4\mathbb{Q}^{2}}}\right)}{2\sqrt{-1+4\mathbb{Q}^{2}}},
\end{equation}
this correspond to (\ref{eq:QVariance}) upto $\mathrm{\gamma}$.

\subsection{Holography}

Assume now that a flat spacetime with an additional coarse-grained
direction $u$ is seen as an AdS space in $C_{T}\to0$, and that the
metric is of the form

\begin{equation}
ds^{2}=g_{uu}\epsilon^{2}\left(du^{2}+dx^{2}-dt^{2}\right),
\end{equation}

\begin{equation}
v\equiv\frac{1}{\sqrt{L_{T}C_{T}}}.
\end{equation}

We obtaon $g_{uu}$ from $|\Phi\rangle$, 
\begin{equation}
\begin{aligned}g_{uu}(u) & =\langle\Psi_{I}(u)|K_{I}(u)^{2}|\Psi_{I}(u)\rangle-\langle\Psi_{I}(u)|K_{I}(u)|\Psi_{I}(u)\rangle^{2}\\
 & =\chi(u)^{2}=\frac{1}{4}\left(\frac{1}{1+\frac{L_{T}}{L}\frac{1}{\Lambda e^{-2u}}}\right)^{2}=\frac{1}{4}\left(\frac{1}{1+\frac{L_{T}}{L}\frac{\epsilon^{2}}{z^{2}\Lambda}}\right)^{2}.
\end{aligned}
\end{equation}

where we introduce $z$ as 
\begin{equation}
z=\epsilon e^{-u},\ \frac{\epsilon}{z}=e^{u}.
\end{equation}

we obtain the metric: 
\begin{equation}
ds^{2}=\frac{1}{4}\frac{1}{z^{2}}\left(\frac{1}{1+\frac{L_{T}}{L\Lambda}\frac{\epsilon^{2}}{z^{2}}}\right)^{2}\left(dz^{2}+dx^{2}-dt^{2}\right).\label{eq:endpoint_metric}
\end{equation}

In $\frac{L_{T}}{L\Lambda}\frac{\epsilon^{2}}{z^{2}}\ll1,$ is almost
$\mathrm{AdS}_{3}$.

From the metric, the Einstein tensor is

\begin{equation}
G_{\mu\nu}=R_{\mu\nu}-\frac{1}{2}g_{\mu\nu}R,
\end{equation}
where $R_{\mu\nu}$ is Ricci tensor and $R$ is Ricci tensor:

\begin{equation}
\begin{aligned}R_{\mu\nu} & =-\frac{2}{z^{2}\left(1+\frac{L_{T}}{L}\frac{z^{2}}{\epsilon^{2}\Lambda}\right)}\mathrm{diag}\left(1+\frac{L_{T}}{L}\frac{z^{2}}{\epsilon^{2}\Lambda}+2\left(\frac{L_{T}}{L}\frac{z^{2}}{\epsilon^{2}\Lambda}\right)^{2},\right.\\
 & ,\left.1+\frac{5}{2}\frac{L_{T}}{L}\frac{z^{2}}{\epsilon^{2}\Lambda}+3\left(\frac{L_{T}}{L}\frac{z^{2}}{\epsilon^{2}\Lambda}\right)^{2},1+\frac{5}{2}\frac{L_{T}}{L}\frac{z^{2}}{\epsilon^{2}\Lambda}+3\left(\frac{L_{T}}{L}\frac{z^{2}}{\epsilon^{2}\Lambda}\right)^{2}\right)\\
R & -64\frac{L_{T}}{L}\frac{z^{2}}{\epsilon^{2}\Lambda}-40\frac{1}{1+\frac{L_{T}}{L}\frac{z^{2}}{\epsilon^{2}\Lambda}}.
\end{aligned}
,
\end{equation}

Contrary to the usual theory of gravity, let us now try to find the
energy-momentum tensor under the AdS/CFT correspondence, assuming
that the metric obtained from the state obtained by cMera satisfies
the Einstein equation.

Introducing the cosmological term $\lambda$ and energy-momentum term,
Einstein equation is

\begin{equation}
R_{\mu\nu}-\frac{1}{2}g_{\mu\nu}R+\lambda g_{\mu\nu}=\kappa T_{\mu\nu}
\end{equation}
where 
\begin{equation}
\begin{aligned}T_{\mu\nu} & =\frac{1}{z^{2}}\frac{1}{\left(1+\frac{L_{T}}{L}\frac{z^{2}}{\epsilon^{2}\Lambda}\right)^{2}}\mathrm{diag}\left(\lambda+\left(1+2\frac{L_{T}}{L}\frac{z^{2}}{\epsilon\Lambda}\right)^{2},1+\lambda+\frac{L_{T}}{L}\frac{z^{2}}{\epsilon^{2}\Lambda}+2\left(\frac{L_{T}}{L}\frac{z^{2}}{\epsilon^{2}\Lambda}\right)^{2}\right.\\
 & \left.-1-\lambda-\frac{L_{T}}{L}\frac{z^{2}}{\epsilon^{2}\Lambda}-2\left(\frac{L_{T}}{L}\frac{z^{2}}{\epsilon^{2}\Lambda}\right)^{2}\right)
\end{aligned}
\end{equation}

From the continuity formula, $g^{\mu\nu}\nabla_{\nu}T_{\mu\rho}=0$,
the cosmic cosmological $\lambda$ is obtained as

\begin{equation}
\lambda=-\frac{4\left(1+\frac{L_{T}}{L}\frac{z^{2}}{\epsilon^{2}\Lambda^{2}}+2\left(\frac{L_{T}}{L}\frac{z^{2}}{\epsilon^{2}\Lambda^{2}}\right)^{2}\right)}{1+\frac{L_{T}}{L}\frac{z^{2}}{\epsilon^{2}\Lambda^{2}}}.
\end{equation}

In the limit $L_{T}\to0$, $\lambda$ is $-4$. This result is consistent
with the $\mathrm{AdS_{3}}$ metric.

Next, following Witten's approach\cite{Witten_1998}, let us find
the propagator of the boundary and the propagator of the bulk and
discuss the $(L,L_{T})$ dependence of the propagator of the bulk.
We describe this propagator by $K$.

In general, the Laplacian of a three-dimensional massless scalar field
satisfies the following equation:

\begin{equation}
\square K=\frac{1}{\sqrt{|g|}}\partial_{\mu}\left(\sqrt{|g|}g^{\mu\nu}\partial_{\nu}K\right)=0.
\end{equation}

In our case (\ref{eq:endpoint_metric}), the metric is 
\begin{equation}
\{g_{\mu\nu}\}=\frac{1}{4}\frac{1}{z^{2}}\left(\frac{1}{1+\frac{L_{T}}{L}\frac{z^{2}}{\epsilon^{2}\Lambda}}\right)^{2}\left(\begin{array}{ccc}
1 & 0 & 0\\
0 & 1 & 0\\
0 & 0 & -1
\end{array}\right)
\end{equation}

Therfore, the inverce metric and square root of the determinat $\sqrt{|g|}$
are

\begin{equation}
\{g^{\mu\nu}\}=4z^{2}\left(1+\frac{L_{T}}{L}\frac{z^{2}}{\epsilon^{2}\Lambda}\right)^{2}\left(\begin{array}{ccc}
1 & 0 & 0\\
0 & 1 & 0\\
0 & 0 & -1
\end{array}\right),
\end{equation}

\begin{equation}
\sqrt{|g|}=\frac{1}{8}\frac{1}{z^{3}}\left(\frac{1}{1+\frac{L_{T}}{L}\frac{z^{2}}{\epsilon^{2}\Lambda}}\right)^{3}.
\end{equation}

Considering those ignored due to symmetry and translation invariance
except in the $z$-direction, we obtain 
\begin{equation}
\partial_{z}\left(\frac{1}{2}\frac{1}{z}\left(\frac{1}{1+\frac{L_{T}}{L}\frac{z^{2}}{\epsilon^{2}\Lambda}}\right)\partial_{z}K(z)\right)=0
\end{equation}

The solution that vanishes at $x_{0}=0$ is

\begin{equation}
K(z)=c\left(1+\frac{1}{2}\frac{L_{T}}{L}\frac{z^{2}}{\epsilon^{2}\Lambda}\right)z^{2}
\end{equation}
with $c$ being a constant.

The solution takes the form of a singularity at the infinity point.

Using $SO(1,2)$ transformation, we map the infinity point to origin,

\begin{equation}
x^{\mu}\to\frac{x^{\mu}}{z^{2}+x^{2}-t^{2}}.
\end{equation}

Then 
\begin{equation}
K(x^{\mu})=c\left(1+\frac{1}{2}\frac{L_{T}}{L}\frac{z^{2}}{\epsilon^{2}\Lambda}\frac{1}{\left(z^{2}+x^{2}-t^{2}\right)^{2}}\right)\frac{z^{2}}{\left(z^{2}+x^{2}-t^{2}\right)^{2}}.
\end{equation}

We express the boundary field as $\phi_{0}(\text{\ensuremath{\bm{x}}})$
,where $\bm{x}=(x,t)$.

Using this Green's function, $\phi(z,x,t)$ in AdS space is

\begin{equation}
\phi(z,x,t)=c\int d\bm{x'}\left(1+\frac{1}{2}\frac{L_{T}}{L}\frac{z^{2}}{\epsilon^{2}\Lambda}\frac{1}{\left(z^{2}+|\bm{x}-\bm{x}'|^{2}\right)^{2}}\right)\frac{z^{2}}{\left(z^{2}+|\bm{x}-\bm{x}'|^{2}\right)^{2}}\phi_{0}(\bm{x}).
\end{equation}

This result show that the bulk to boundary propagator depends on ratio
$\frac{L_{T}}{L}$. This dependency indicates how the properties of
the transmission line, such as the inductance per unit length $L_{T}$
and endpoint inductance $L$, influence the behavior of the field
$\phi$ in the bulk of the holographic space. It also provides a means
to extract physical properties from the boundary conditions.

\section{Conclusion and Discussion}

Since the quantum theory of transmission lines can be regarded as
a one-dimensional Klein-Gordon equation, we have shown a procedure
to calculate the entanglement renormalisation group using the continuous
MERA argument to construct states at the renormalisation scale μ of
the vacuum state and obtain the holographic space-time corresponding
to the state of the transmission line in Section 3.

Then, as a non-trivial example in quantum networks, the entanglement
renormalisation group of a transmission line with endpoints was performed
and the holographic spacetime corresponding to this state was obtained.
It was argued that the metric is obtained in a way that the scale
factor depends on the ratio of the capacities of the transmission
line and the endpoints, and corresponds to the $\mathrm{AdS_{3}}$
spacetime in the limit where the capacitance of the endpoints is zero
in Section 5.

This suggests that superconducting quantum circuits themselves, including
various electric circuits and quantum computers, can be described
in some unified geometry. In particular, it is interesting to see
how modularity, which is unique to electric circuits and superconducting
circuits, can be expressed on the gravitational side. Furthermore,
the renormalisation group view of the network itself, which performs
more complex qubit calculations, may shed new light on issues such
as quantum errors, which have been a problem in quantum computers.

The next step is to deal with more complex quantum networks for connection
to quantum computers (qubits), which will naturally need to deal with
non-linear interactions and strongly coupled cases as interaction
Hamiltonians. Several methods for treating interactions as perturbations
in cMera have been discussed\cite{Bhattacharyya_2018,https://doi.org/10.48550/arxiv.1612.02427,Cotler_2019_E,Cotler_2019_R,FernandezMelgarejo2020},
and in particular, methods using exact renormalisation groups may
be useful in tackling such issues\cite{Kuwahara_2023,Goldman_2023}.
In particular, recent discussions of holography with random tensor
networks\cite{Hayden_2016} are useful in the case of strong coupling
and may be powerful for renormalisation group-like analyses of complex
quantum networks such as the one in this paper.

Such attempts provide a starting point for computing more complex
quantum circuits, and many interesting examples could be discussed
through holography. In particular, it is interesting to see how resonant
states such as laser oscillation correspond to the behaviour of a
theory of gravity. It is also of interest to compare the complexity
as the number of elements in such complex circuit electromagnetic
science with the complexity in quantum computation. Hopefully, these
studies will contribute to future research on quantum networks.

\section*{Acknowledgments}

I particularly appreciate Akio Sugamoto for introducing me to Yurke's
paper\cite{Yurke_1984} at the beginning of this research and for
the subsequent in-depth discussions on this research. I also appreciate
Shiro Komata and Tatsuaki Wada for their careful reading of this paper
and many useful comments. Finally, I would like to thank the members
of the Open University of Japan Field Theory Seminar for their many
suggestions and advice in the progression of this research.

\section*{Appendix A: Mera and AdS}

This section explains how Mera corresponds with AdS\cite{Swingle_2012}.

Mera was designed to find the ground state of a quantum many-body
system at the critical point, i.e. the CFT. $d_{s}$-dimensional quantum
critical systems usually have an area law for the entanglement entropy\cite{Eisert_2010},
which is

\begin{equation}
S_{A}\propto L^{d_{s}-1}
\end{equation}
with respect to the linear length $L$ in the domain $A$ of the system.

However, for one-dimensional quantum critical systems, the entanglement
entropy breaks the area law and depends on the logarithm of the scale
$L$ and is

\begin{equation}
S_{A}=\frac{c}{3}\log L.
\end{equation}

Such a violation of the area law is discussed by introducing the scale
$z$ of the renormalization group. Each scale $z$ has different degrees
of freedom. At more coarse-grained scales, the degrees of freedom
visible at finer scales are grouped together and treated as a single
degree of freedom. Consider these degrees of freedom on a logarithmic
scale $d\log z$. The degrees of freedom at each scale can be entangled
with a region $A$ with linear size $L$ and contribute to the entropy
proportional to the boundary $\partial A$ of region $A$ at that
scale $z$. The number of degrees of freedom localized at $\partial A$
is simply proportional to the logarithmic scale $d\log z$, and the
entanglement entropy of region $A$ at scale $z$ is proportional
to the degrees of freedom of its boundary by the area law, so

\begin{equation}
dS_{A}\sim\frac{L^{d_{s}-1}dz}{z^{d_{s}-1}}.
\end{equation}

In $d_{s}=1$, the overall entanglement entropy is

\begin{equation}
S_{A}\sim\int_{a}^{\xi_{E}}\frac{dz}{z}=\log\left(\frac{\xi_{E}}{a}\right)
\end{equation}
where $a$ is ultraviolet cutoff and $\xi_{E}$ is a length scale
such that there is no entanglement in the quantum state and it diverges
at the quantum critical point. However, in reality, there is a finite
size $L$ in region $A$, so a cut-off is introduced and the entropy
is

\begin{equation}
S_{A}\sim\log\left(\frac{L}{a}\right).
\end{equation}

In this way, the breaking of the area law in one-dimensional critical
quantum many-body systems can be explained from the perspective of
the renormalisation group.

Since it is known from theoretical and numerical arguments based on
scale invariance and fractality of the system at the quantum critical
point that the entanglement entropy is proportional to the length
of the minimum curve $\gamma$ in region $A$\cite{Pfeifer_2009},
the following equation is obtained

\begin{equation}
ds_{\gamma}^{2}=dS_{A}^{2}\sim\frac{dz^{2}}{z^{2}}.
\end{equation}

In this way, the corresponding part of the scale transformation in
the AdS space is reproduced. Other parts of the AdS space are unknown.

\section*{Appendix B: Dimensional analysis for transmission lines}

This section summarizes the dimensional analysis for transmission lines.
Because \(H\) is energy density, we analyze the dimensional consistency of the expressions. First, consider the expression for the relationship between mass, displacement, and time:
\begin{align} \frac{[m x]}{[t^2]} &= \frac{[\Phi^2]}{[L]} = [Q^2][C x^2], \label{eq:dim1}  \end{align}
Here, \(\Phi\) represents the canonical conjugate of the charge \(Q\) in the quantization of the LC circuit. The relationship between \(\Phi\) and \(Q\) can be expressed as:
\begin{align} \Phi &= L \frac{\partial Q}{\partial t}, \quad \text{we obtain:} \\ [\Phi] &= \frac{[L Q]}{[t]}, \label{eq:dim2} \end{align}
Next, consider the dimensional relation involving position and time:
\begin{align} \frac{[x^2]}{[t^2]} &= \frac{1}{[L_E C]}, \label{eq:dim3} \end{align}
For the capacitance \(C_T\) and inductance \(L_T\), the dimensional consistency is given by:
\begin{align} [C_T] &= \frac{[Q^2 t^2]}{[m x^3]} = \frac{[Q^2]}{[E x]}, \label{eq:dim4} \\ [L_T] &= \frac{[m x]}{[Q]^2} = \frac{[E t^2]}{[Q^2 x]}, \label{eq:dim5} \end{align}
Finally, the dimensional consistency for the canonical conjugate \(\Phi\) and charge \(Q\) is:
\begin{align} [\Phi] &= \frac{[m x]}{[t Q]} = \frac{[E t]}{[Q x]}, \label{eq:dim6} \end{align}
Now, let's consider the case where \(Q(t, x)\) is the charge density, which depends on both time \(t\) and position \(x\). When performing a Fourier transform with respect to \(x\), the resulting function \(Q(t, k)\) depends on the wavevector \(k\), where \(k\) is conjugate to the position \(x\). Similarly, \(\Phi(t, k)\) is the Fourier transform of \(\Phi(t, x)\). To maintain dimensional consistency after the Fourier transform, the relationships are:
\begin{align} [Q(t, k)] &= [x Q(t, x)], \label{eq:dim7} \\ [\Phi(t, k)] &= [x \Phi(t, x)]. \label{eq:dim8} \end{align}

\bibliographystyle{unsrt}
\bibliography{trasmissionLine_paper_latagiri}

\end{document}